\documentclass{PoS}
\usepackage{amsmath,amssymb}

\title{Study of the pion-pion scatterings with a combination of all-to-all propagators and the HAL QCD method}

\author{\speaker{Yutaro Akahoshi}$^{1,2}$, Sinya Aoki$^{1,2}$, Tatsumi Aoyama$^{2,3}$, Takaya Miyamoto$^{2}$, and Kenji Sasaki$^{1,2}$, for HAL QCD Collaboration\\
        \llap{$^1$}Center for Gravitational Physics, Yukawa Institute for Theoretical Physics, Kyoto University, Kyoto 606-8502, Japan \\
        \llap{$^2$}RIKEN Nishina Center(RNC), Saitama 351-0198, Japan \\
        \llap{$^3$}Institute of Particle and Nuclear Studies, High Energy Accelerator Research Organization(KEK), Tsukuba, Ibaraki 305-0801, Japan \\
        E-mail: \email{yutaro.akahoshi@yukawa.kyoto-u.ac.jp},\\
        \email{saoki@yukawa.kyoto-u.ac.jp},\\ \email{aoym@post.kek.jp},\\
        \email{miyamoto@ribf.riken.jp},\\
        \email{kenjis@yukawa.kyoto-u.ac.jp}}

\abstract{
In this paper, we report recent developments of the HAL QCD method
for two hadron systems which contain quark annihilation processes
using all-to-all quark propagators.
We employ the hybrid method for all-to-all propagators, which combines a low-mode spectral
decomposition of the quark propagator and stochastic estimators for remaining
high modes, to evaluate the HAL QCD potentials for the first time.
Using this method, we investigate the $I=
1,2$ $\pi \pi$ scatterings at $m_{\pi} \approx 870$ MeV.
In the $I=2$ study, we study how statistical fluctuations of the HAL QCD potentials
are increased due to stochastic estimators in the hybrid method, compared with the conventional one
without them.
We find that we can reduce statistical fluctuations by dilutions of stochastic noises in
order to obtain sufficiently precise results, which turn out to be consistent with conventional results without all-to-all propagators.
In the $I=1$ $\pi \pi$ case, which contains quark annihilation processes, we find that
statistical fluctuations are further enhanced due to noise contaminations in annihilation processes. We, however, confirm that we can also reduce such statistical fluctuations to obtain
the potential with a reasonable precision as long as
we further increase a degree of dilutions at a price of large numerical costs and take an appropriate scheme for the potential.
}

\FullConference{37th International Symposium on Lattice Field Theory - Lattice2019\\
		16-22 June 2019\\
		Wuhan, China}

\ShortTitle{Study of the $\pi \pi$ scatterings with all-to-all propagators and HAL QCD method}

\begin{document}

\section{Introduction}
It has been observed that there are some unconventional hadronic resonances, such as $X, Y, Z$ states and $\sigma$ meson. Those states have been widely studied by using some models, but their properties are still not well-understood.
Therefore, model-independent investigation of exotic states by lattice QCD is strongly needed to understand them more accurately.

The HAL QCD method\cite{Ishii:2006ec,Aoki:2009ji,Aoki:2011ep,HALQCD:2012aa} is a way to construct inter-hadron potentials from lattice QCD with a strong advantage that it enables us to study pole structures of the S-matrix directly without any model-dependent ansatz. Therefore, the analysis by using the HAL QCD method is mandatory to understand the nature of hadronic resonances.
To study hadronic resonances in the HAL QCD method
generally requires  all-to-all propagators within reasonable numerical costs,
which has not been established yet for the potentials.
Under such circumstances, we have been testing the hybrid method\cite{Foley:2005ac},
which evaluates all-to-all quark propagators by the spectral decomposition for the low-mode part plus the stochastic estimation for the high-mode part.
In this paper, we will report two results,  the $I=2$ S-wave $\pi \pi$ scattering study\cite{Akahoshi:2019klc}, which is intended to study how the statistical errors of the potential behaves with the hybrid method, and the $I=1$ P-wave $\pi \pi$ scattering study, which is a preparatory research for future $\rho$ resonance studies. From these results, we confirmed that the hybrid method works well in the HAL QCD method as far as we reduce additional statistical fluctuations due to the stochastic estimation in the hybrid method.

\section{HAL QCD method}
The fundamental quantity in the HAL QCD method is the Nambu--Bethe--Salpeter (NBS) wave function, which is defined as
\begin{equation}
  \psi_{W}({\bf r},\Delta t) = \langle 0| \pi({\bf x+r},\Delta t) \pi({\bf x},0) |\pi \pi;{\bf k} \rangle,
\end{equation}
where $\pi({\bf x},t)$ is the pion operator, and
$|\pi \pi;{\bf k} \rangle$ is the elastic state of the two-pion system with a relative momentum ${\bf k}$. We here introduce a relative time difference $\Delta t$ between two pion operators at the sink,
while  $\Delta t = 0$ has been exclusively used in the previous HAL QCD studies.
In this study we take local operators for positively(negatively)-charged pion, $\pi^{+}({\bf x},t) = \bar d({\bf x},t) \gamma_5 u({\bf x},t)$ ($\pi^{-}({\bf x},t) = \bar u({\bf x},t) \gamma_5 d({\bf x},t)$).

We extract the potential from the normalized correlator $R({\bf r},t,\Delta t)$, which is a sum of NBS wave functions as
\begin{equation}
  R({\bf r},t,\Delta t) \equiv \frac{\langle 0| \pi({\bf x+r},t+\Delta t) \pi({\bf x},t) {\mathcal J}_{\pi \pi} (0)|0 \rangle}{C_{\pi \pi}(t)^2} \approx \sum_n A_n \psi_{W_n}({\bf r},\Delta t) e^{-(W_n - 2m_{\pi}) t} + ...\ ,
\end{equation}
where $C_{\pi \pi}$ is the pion two-point function, ${\mathcal J}_{\pi \pi}$ is a source operator which creates $\pi \pi$ states, $W_n$ is the energy of the $n$th elastic state and an ellipsis indicates inelastic contributions.  By using an asymptotic behavior of the NBS wave function\cite{Aoki:2009ji}, we can define a non-local potential as\cite{HALQCD:2012aa}
\begin{equation}
  \left[ \frac{\nabla^2}{m_{\pi}} -\frac{\partial}{\partial t} + \frac{1}{4m_{\pi}} \frac{\partial^2}{\partial t^2} \right] R({\bf r},t,\Delta t) = \int d^3{\bf r'} U({\bf r},{\bf r'}) R({\bf r'},t,\Delta t).
\end{equation}
In practice, the non-locality of the potential is treated by the derivative expansion,
\begin{equation}
  U({\bf r},{\bf r'}) = (V_0(r) + V_1(r) \nabla^2 + {\mathcal O}(\nabla^4)) \delta({\bf r-r'}),
\end{equation}
whose leading-order (LO)  term, for example, is given by
\begin{equation}
  V^{\rm LO}(r) = \frac{\left[ \frac{\nabla^2}{m_{\pi}} -\frac{\partial}{\partial t} + \frac{1}{4m_{\pi}} \frac{\partial^2}{\partial t^2} \right] R({\bf r},t,\Delta t)}{R({\bf r},t,\Delta t)}.
\end{equation}

\section{The hybrid method for all-to-all propagators}
In this study, we employ the hybrid method\cite{Foley:2005ac}, which enables us to obtain
all-to-all propagators while keeping the locality of the quark operators. The starting point is the low-mode spectral decomposition of the quark propagator,
\begin{equation}
  D_0^{-1}(x,y) = \sum_{i=0}^{N_{\rm eig}-1} \frac{1}{\lambda_i} v^{(i)}(x) \otimes v^{(i)}(y)^{\dag} \gamma_5,
\end{equation}
where $\lambda_i, v^{(i)}$ are the $i$th eigenvalues and eigenvectors of $H=\gamma_5 D$, respectively, and $N_{\rm eig}$ is a number of low-modes we take in the calculation.
The remaining high-mode part is estimated stochastically using the noise vector $\eta_{[r]}^{(i)}$, together with the noise reduction by dilution technique\cite{Foley:2005ac} as
\begin{equation}
  D^{-1} - D_0^{-1} = H^{-1} {\mathcal P}_1 \gamma_5 \approx \frac{1}{N_{\rm r}} \sum_{r=0}^{N_{\rm r}-1} \sum_{i=0}^{N_{\rm dil}-1} \psi_{[r]}^{(i)}(x) \otimes \eta_{[r]}^{(i)}(y)^{\dag} \gamma_5,
\end{equation}
where ${\mathcal P}_1 \equiv {\bf 1} - \sum_{i=0}^{N_{\rm eig}-1} v^{(i)} \otimes v^{\dag(i)}$ is a projection to the high-eigenmode space, $N_{\rm r}$ ($N_{\rm dil}$) is a number of noise vectors (dilutions), and $\psi_{[r]}^{(i)}$ are solution to $ H \cdot \psi_{[r]}^{(i)}= {\mathcal P}_1 \eta_{[r]}^{(i)}$.

\vspace{1mm}
In summary, we can write the quark propagator in the hybrid method as
\begin{equation}
  D^{-1} \approx \frac{1}{N_{\rm r}} \sum_{r=0}^{N_{\rm r}-1} \sum_{i=0}^{N_{\rm hl}-1} u_{[r]}^{(i)} \otimes w_{[r]}^{\dag (i)} \gamma_5,\quad
 N_{\rm hl} = N_{\rm eig} + N_{\rm dil},
\end{equation}
where $u$ and $w$ are defined as
\begin{eqnarray}
  w_{[r]}^{(i)} &=& \{ \frac{v^{(0)}}{\lambda_0}, \cdots , \frac{v^{(N_{\rm eig}-1)}}{\lambda_{N_{\rm eig}-1}},\eta_{[r]}^{(0)}, \cdots ,\eta_{[r]}^{(N_{\rm dil}-1)} \} \\
  u_{[r]}^{(i)} &=& \{ v^{(0)}, \cdots , v^{(N_{\rm eig}-1)}, \psi_{[r]}^{(0)}, \cdots ,\psi_{[r]}^{(N_{\rm dil}-1)} \} .
\end{eqnarray}

\section{$I=2$ $\pi \pi$ S-wave scattering}
We first study the $I=2$ $\pi \pi$ S-wave scattering, using
2+1 flavor full QCD configurations on $16^3 \times 32$ lattice generated by the CP-PACS and JLQCD Collaborations\cite{Ishikawa:2007nn} with the Iwasaki gauge action\cite{Iwasaki:1985we} and a non-perturbatively ${\mathcal O}(a)$ improved Wilson-Clover action\cite{Sheikholeslami:1985ij}. In this setup, the lattice spacing is $a \approx 0.1214$ fm and the pion mass is $m_{\pi} \approx 870$ MeV.
For the NBS wave function, we take the equal--time scheme($\Delta t = 0$).
We employ the smeared quark source $q_s({\bf x},t) = \sum_{\bf y} f({\bf x-y}) q({\bf y},t)$,
together with the Coulomb gauge fixing,
to achieve the ground state saturation at early imaginary times,
where
$f( {\bf x}) = \{a e^{-b |{\bf x }|}, 1, 0\}$ for $\{0 < |{\bf x }| < (L-1)/2, |{\bf x }| = 0, |{\bf x }| \geq (L-1)/2\}$ with  $a=1.0,b=0.47$ in lattice units.
For the hybrid method, we take $N_{\rm eig} = 100$ and $n_r=1$ for $Z_4$ noises, together with full dilutions for color and Dirac indices, the 16-interlace dilution for time and $s4$ dilution for space.
(Thus $N_{\rm dil} = 3\times 4\times 16\times 4$. See Ref.~\cite{Akahoshi:2019klc} for more details.)
Statistical errors from $N_{\rm conf} = 60$ are estimated by the jackknife method with bin--size 6.
We use the wall source result at $t=10$  without all-to-all propagators for a comparison.

\begin{figure}[tbp]
  \hspace{-10mm}
  \centering
  \begin{tabular}{cc}
  \begin{minipage}{0.5\hsize}
    \includegraphics[width=\hsize,bb=0 0 798 574,clip]{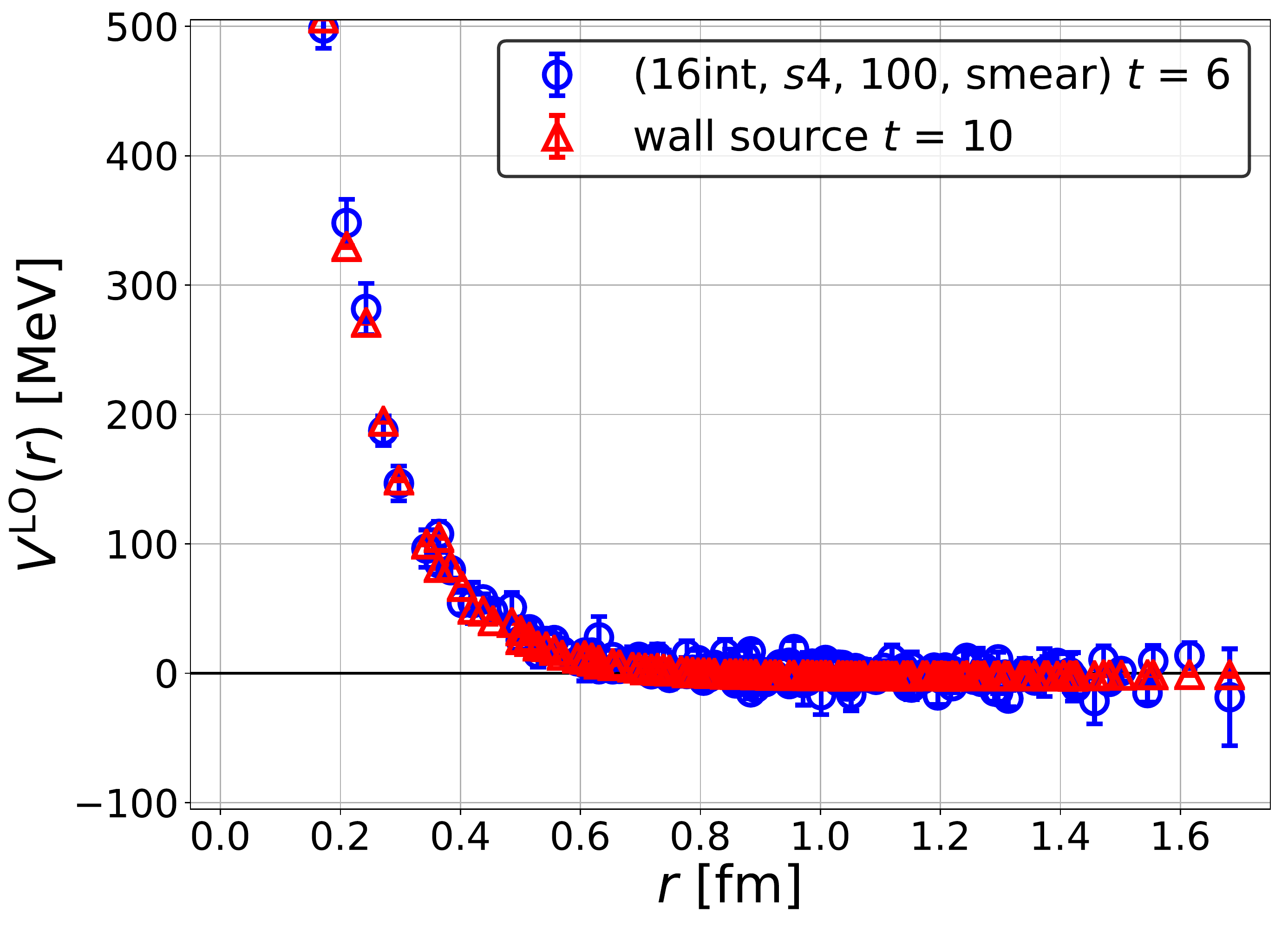}
  \end{minipage} &
  \begin{minipage}{0.5\hsize}
    \includegraphics[width=\hsize,bb=0 0 798 574,clip]{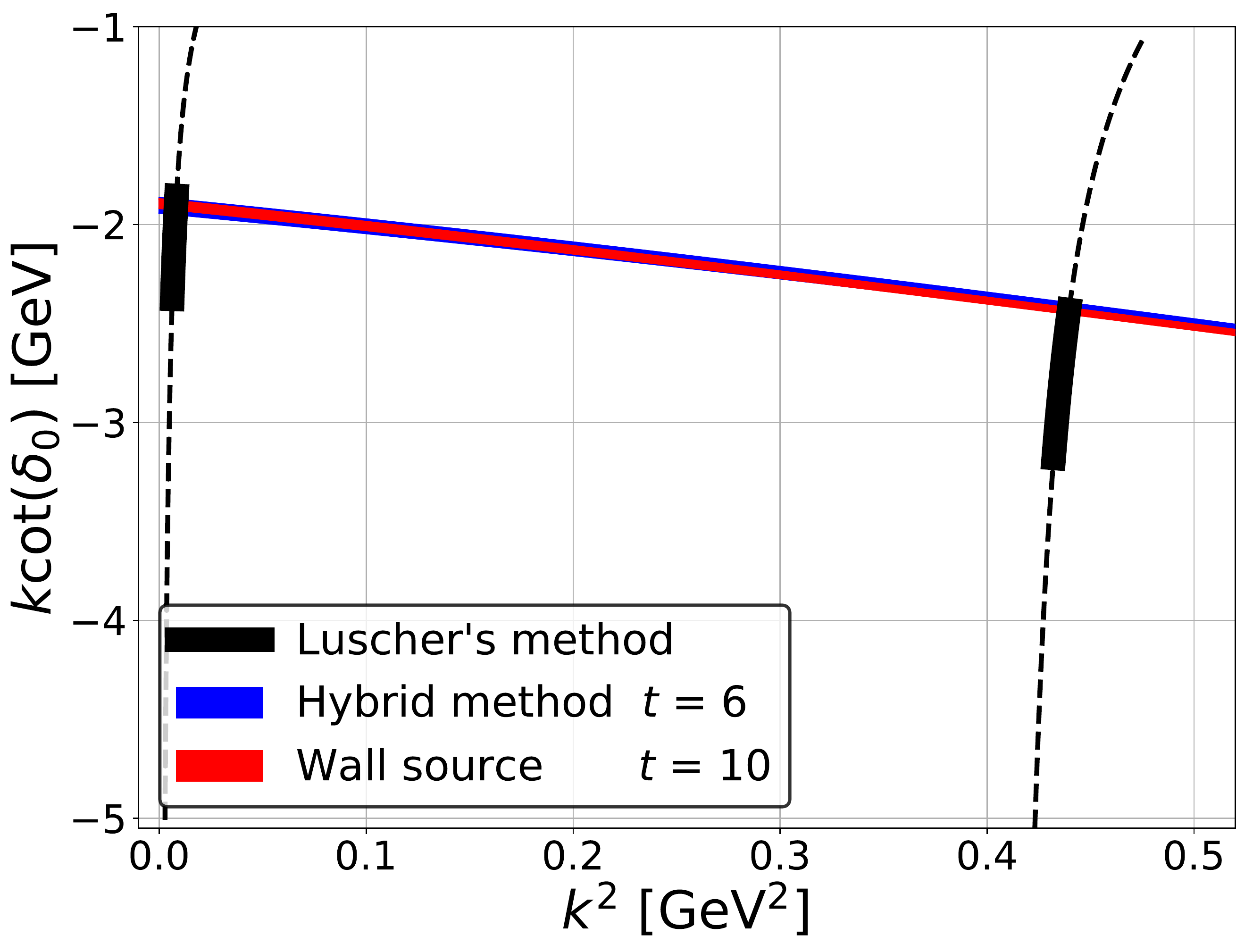}
  \end{minipage}
\end{tabular}
\caption{(Left) The LO potential by the hybrid method (blue) and by the conventional method with the wall source (red). (Right) $k \cot \delta_0$ obtained from the hybrid method (blue) and the conventional one with the wall source (red), together with the result of L\"uscher's method(black).}
\label{fig:i2pp_potential}
\end{figure}

The LO potential from the hybrid method is compared with the one from the wall source in Fig.\ref{fig:i2pp_potential}(left).
Although the statistical fluctuations are somewhat enhanced due to the additional stochastic noises in the hybrid method, both results agree with each other
\footnote{Generally, the LO potentials from different source operators are not identical due to the higher order terms in the derivative expansion.
The agreement suggests either that higher order terms have negligible contributions or that
both correlators are dominated by the ground state.}.
Note that to achieve such high precision with the hybrid method, it is very important to take fine dilutions in spatial directions.
The potential is fitted by the sum of two Gaussian functions,
\begin{equation} \label{eq:i2pp_fitfunc}
 V(r) = \sum_{i=0}^{N-1} a_{2i} e^{-(\frac{r}{a_{2i+1}})^2}
\end{equation}
with $N=2$.
We solve Schr\"odinger equation with the fitted potential, to extract  $k \cot \delta_0(k)$,
which are shown in Fig.\ref{fig:i2pp_potential}(Right), together with the result from the L\"uscher's method.
As expected from the agreement of the potentials, the phase shifts are consistent with each other, and also agree with the result from the L\"uscher's method.
We thus confirm that physical observables can be obtained
with sufficient accuracy in the HAL QCD method combined with all-to-all propagators
from the hybrid method.

We have also systematically studied how the statistical fluctuations of the potentials are affected by the choice of the parameters of the hybrid method.
For more details, see \cite{Akahoshi:2019klc}.

\begin{table}[tbp]
  \hspace{-10mm}
  \caption{Details of setups of the results in $I=1$ $\pi \pi$ study}
  \vspace{2mm}
  \begin{tabular}{c|cccccc}\hline
    & Source & Scheme & time dilution & space dilution & $N_{\rm eig}$ & $N_{\rm conf}$\\ \hline \hline
    case 0 & point & $\Delta t = 0$ & 16-interlace & $s2$ & 100 & 20 \\ \hline
    case 1 & smear & $\Delta t = 1$ &
    \begin{tabular}{l}
      16-intelace (src-to-sink)\\
      4-interlace (sink-to-sink)
    \end{tabular} &
    \begin{tabular}{l}
      $s4$ (src-to-sink)\\
      $s8 \times s2$ (sink-to-sink)
    \end{tabular} & 100 & 60 \\ \hline
  \end{tabular}
  \label{tab:setups_i1pp}
\end{table}
\section{Test calculation for $I=1$ $\pi \pi$ P-wave scattering}
We next study the $I=1$ $\pi \pi$ P-wave scattering, to see how statistical noises in the hybrid method increase if the system contains quark annihilation processes,
using the same gauge ensemble,
where the $\rho$ meson appears, not as a resonance, but as an extremely deep bound state
below the two-pion threshold with  $E_{\rm bind} \approx 510$ MeV.
The setup of our calculations is summarized in Tab.\ref{tab:setups_i1pp}
\hspace{-10mm}
\subsection{$I=1$ $\pi \pi$ potential with the hybrid method}
\begin{figure}[tbp]
  \hspace{-10mm}
  \centering
  \begin{tabular}{cc}
  \begin{minipage}{0.5\hsize}
    \includegraphics[width=\hsize,bb=0 0 827 576,clip]{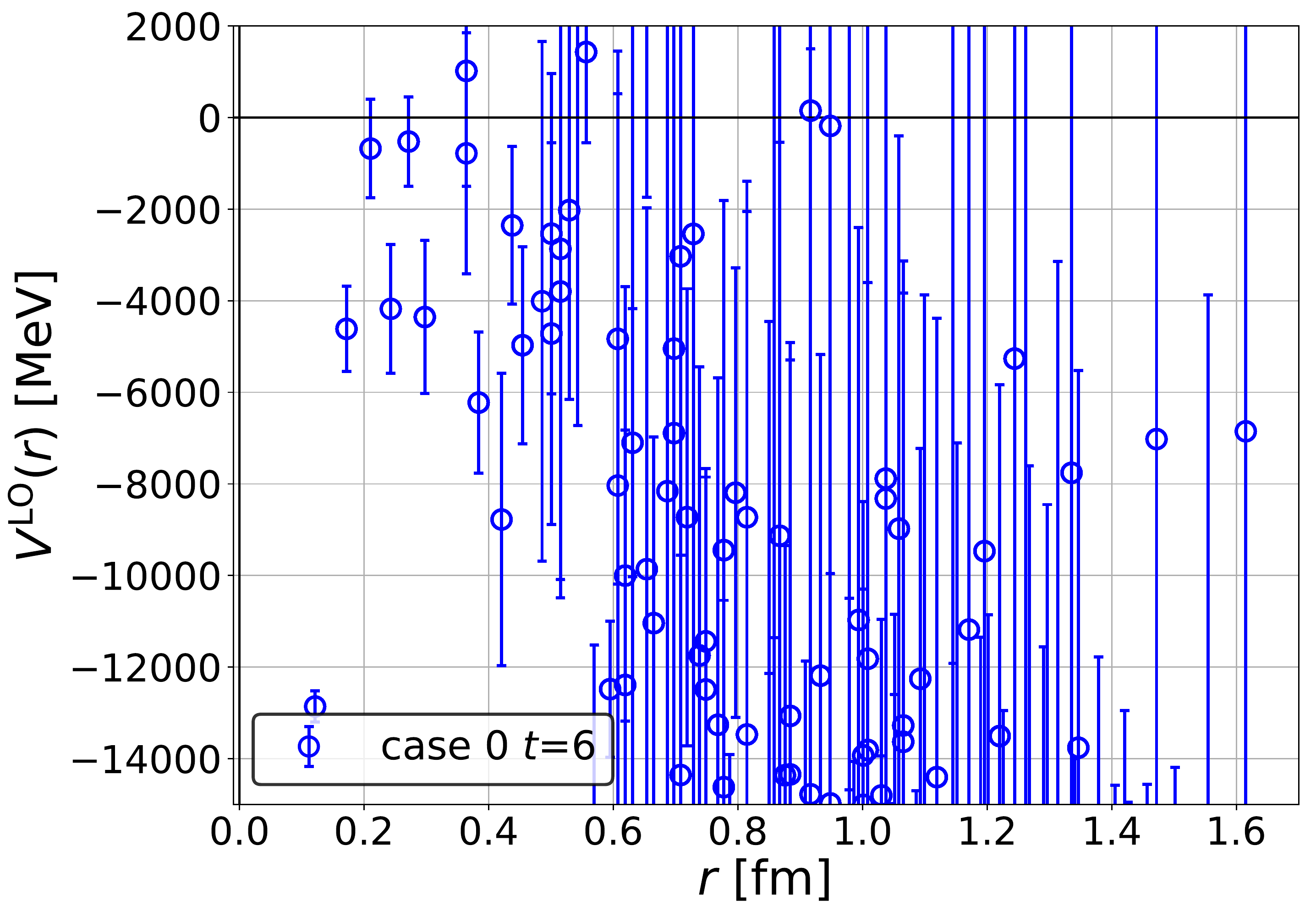}
  \end{minipage} &
  \begin{minipage}{0.5\hsize}
    \includegraphics[width=\hsize,bb=0 0 827 576,clip]{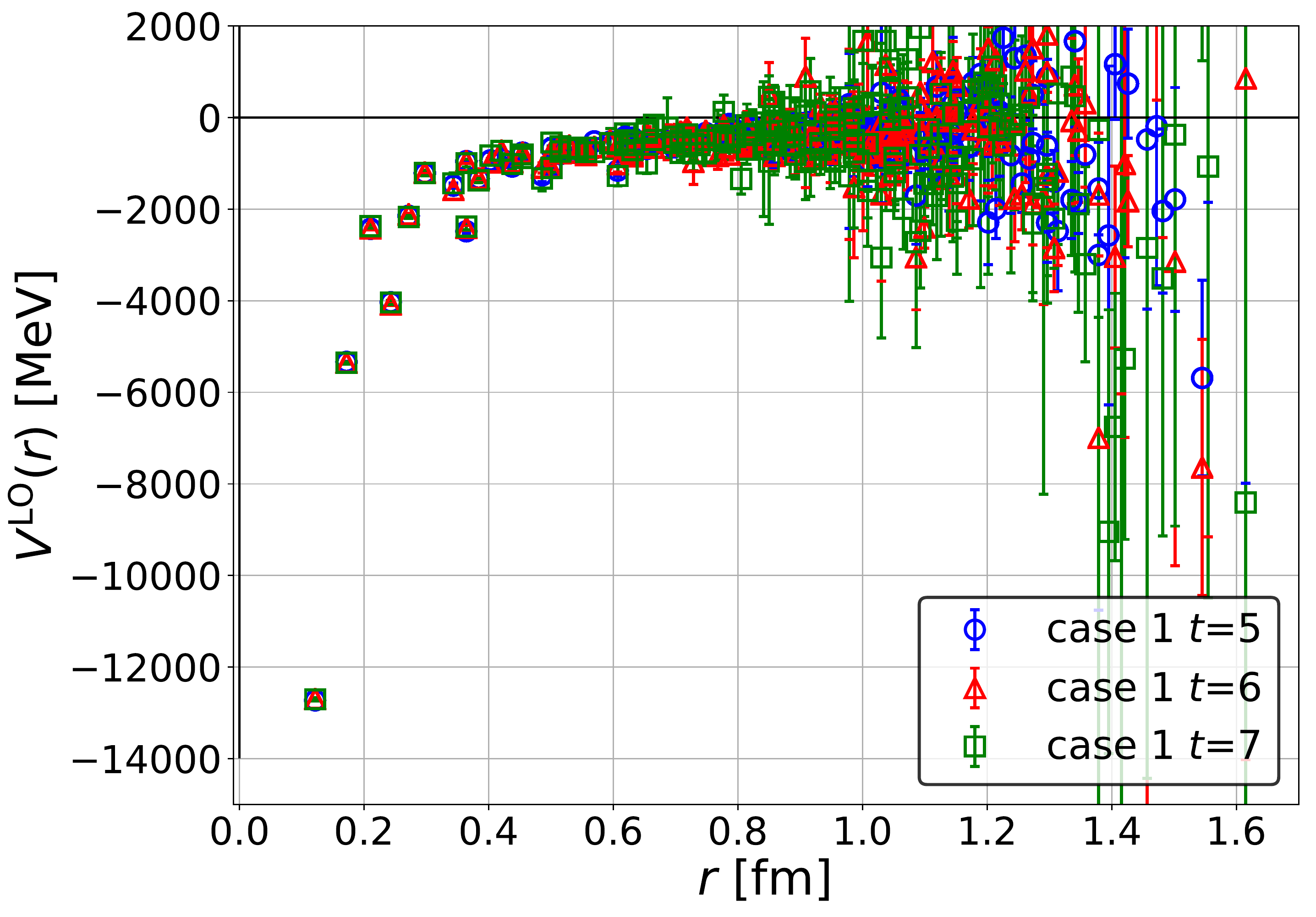}
  \end{minipage}
\end{tabular}
\vspace{-5mm}
\caption{(Left) $I=1$ $\pi \pi$ potential with the same setup in Ref.~\cite{Akahoshi:2019klc}. (Right) The potential with additional noise reductions and its time dependence.}
\label{fig:i1pp_potentials1}
\vspace{-3mm}
\end{figure}

We first employ the setup of the case0 in Tab.\ref{tab:setups_i1pp},
same to the ``case3'' in Ref.~\cite{Akahoshi:2019klc},
which gave the sufficiently precise $I=2$ $\pi \pi$ potential.
The result of the $I=1$ $\pi \pi$ potential is shown in Figure \ref{fig:i1pp_potentials1}(left),
where we observe extremely large statistical fluctuations.
We suspect that
extremely large statistical fluctuations for the $I=1$ $\pi \pi$ potential
are caused by noise contaminations to evaluate equal--time propagations at the sink
in quark annihilation diagrams, which are absent for the $I=2$ $\pi \pi$ potential
in Ref.~\cite{Akahoshi:2019klc}.
Thus, in order to reduce such noise contamination, we employ three additional noise reductions:
(1) The different--time scheme for the NBS wave functions (we take $\Delta t = 1$ in Lattice unit) to avoid the equal--time propagations,
(2) The finer space dilution in the quark annihilation part to reduce noise contamination in spatial indices (8 times finer than case0) and
(3) The average over different noise samples (we take the average over 6 noise samples).
In Fig.\ref{fig:i1pp_potentials1}(right), we show the potential obtained with these noise reductions.
As we can see, the statistical fluctuations of the potential are drastically reduced, and it shows a strong attraction without repulsive core, which is consistent with the existence of the $\rho$ bound state. Note that the potential is almost time-independent, as seen in Fig.\ref{fig:i1pp_potentials1}(right), probably due to the ground state saturation achieved at $t=6$ thanks to the source smearing.
From this result, we establish that the HAL QCD potential can be calculated with sufficient precision by the hybrid method even if the target system contains quark annihilation processes.

\subsection{Binding energy of the $\rho$ meson}
\begin{figure}
  \vspace{-2mm}
  \centering
  \includegraphics[width=0.5\hsize,bb=0 0 843 570,clip]{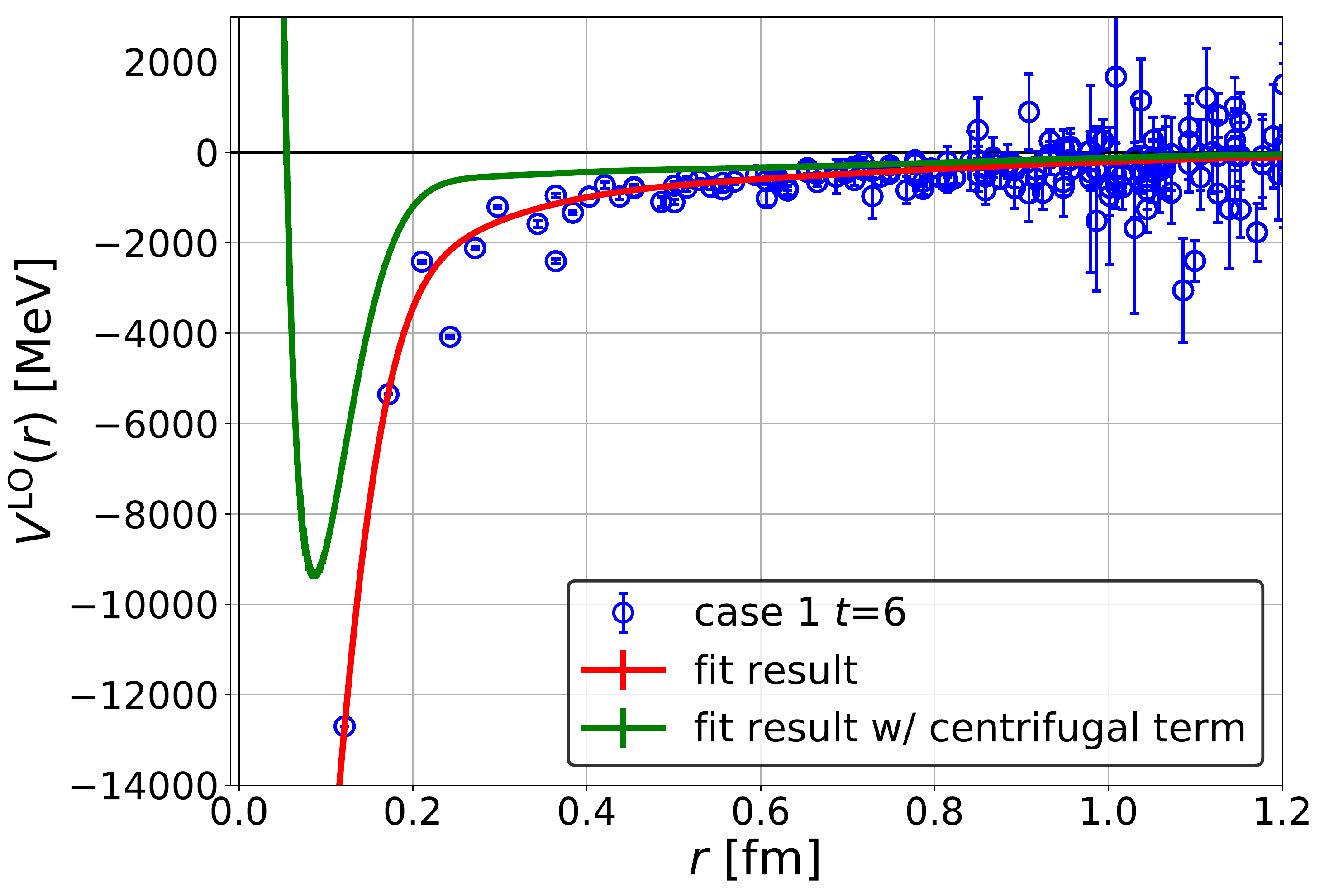}
  \vspace{-3mm}
  \caption{Fitting result of the potential. In this figure, we show $V({\bf r})$ in Eq.\ref{eq:pot_PBC} as a red line. We also show the fitting result with centrifugal term, $V_c(r) = \frac{l (l+1)}{m_{\pi} r^2}$ with $l = 1$(green line).}
  \label{fig:i1pp_pot_fit}
  \vspace{-3mm}
\end{figure}
We evaluate the physical observables, namely the binding energy of the $\rho$ state, by using the $I=1$ $\pi\pi$ potential. For the fitting function, we use the multi-Gaussian shape
(Eq. ~\ref{eq:i2pp_fitfunc}) with $N=3$, and we employ the Gaussian expansion method\cite{Hiyama:2003cu} to calculate the binding energy. The fitting result is shown in Fig.\ref{fig:i1pp_pot_fit}.
Since the potential is slightly deviate from zero at $r = La/2 = 0.9712$ fm,
we include the finite volume effect  from nearest-neighbors into the fitting function as
\begin{equation}
  V({\bf r})_{\rm PBC} = V({\bf r}) + \sum_{{\bf n} \in \{(0,0,\pm1),(0,\pm1,0),(\pm1,0,0)\} } V({\bf r} + L {\bf n}).
  \label{eq:pot_PBC}
\end{equation}
We also exclude some data points which is largely deviates from other data points at $r =$ 0.1214, 0.2428 fm in the fit, since these points are probably caused by the higher partial wave contamination ($l=3$ in this case), which is sometimes observed in the previous HAL QCD studies.

As a result, we obtain the binding energy,
\begin{equation}
  E_{\rm bind} = 668 \pm 24_{\rm stat} \left(\begin{tabular}{c} {\small +69} \\ {\small -151} \end{tabular} \right)_{\rm sys} \ {\rm MeV},
\end{equation}
where
the systematic error is estimated by the time dependence of the binding energy. The central value is somewhat larger than expectation from the single-hadron spectra, $E_{\rm bind, single} \approx 510$ MeV, and it has large time dependence, though the potentials seem to be time-independent. The possible origin of the large time dependence is the systematic uncertainty of the potential fitting at short distances.  As you can see in Fig.~\ref{fig:i1pp_potentials1} (Right), data points in the short distance region are scattered with small statistical errors, and also a number of data points in this region is small. In such a situation, the fit in this region becomes unstable, and as a result, the binding energies strongly depend on time since they are very sensitive to the structure at short distances. To overcome this ambiguity, we have to go finer lattice spacing or
to invent a new scheme of the potential which reduce scatterings of data at short distances.
We leave this problem for future studies.

\section{Summary and outlook}
We apply the hybrid method for all-to-all propagators to
calculations of the HAL QCD potential for $\pi \pi$ systems.
In the $I=2$ case, we confirmed that we can obtain sufficiently accurate results with the hybrid method. In the $I=1$ case, while we observed that the quark annihilation processes enhance
statistical fluctuations due to noise contaminations, we can reduce them to
obtain a potential with reasonable precision even in such a case thanks to additional noise reductions.

Finally, I would like to mention further improvements.
Although we confirm the hybrid method works well with the HAL QCD method, it is also revealed that the numerical cost for the noise reductions is too large to employ simulations with larger lattice sizes.
Therefore, to study hadronic resonances in more physical setups (larger volume, lighter pion, and finer lattice spacing, etc.), further improvement to our calculation scheme is mandatory. Fortunately, we find that by combining some techniques such as the covariant approximation averaging\cite{Shintani:2014vja}, the one-end trick and sequential propagators\cite{Abdel-Rehim:2017dok}, we can achieve both small noise contamination and small numerical cost. Results by using these new techniques will be reported in near future.

\section{Acknowledgements}
The authors thank members of the HAL QCD Collaboration for fruitful discussions. We thank the JLQCD and CP-PACS Collaborations~\cite{Ishikawa:2007nn} and ILDG/JLDG~\cite{Amagasa:2015zwb} for providing their configurations. Numerical simulations are performed on Cray XC40 at YITP in Kyoto University and HOKUSAI Big-Waterfall in RIKEN. The framework of our numerical codes is based on Bridge++ codeset~\cite{Ueda:2014rya}.

\end{document}